\documentclass[conference]{IEEEtran}
\IEEEoverridecommandlockouts
\usepackage{cite}
\usepackage{amsmath,amssymb,amsfonts}
\usepackage{algorithm, algorithmicx}
\usepackage{graphicx}
\usepackage{textcomp}
\usepackage{xcolor}
\usepackage{hyperref}
\usepackage{mathtools}

\usepackage{balance}

\def\BibTeX{{\rm B\kern-.05em{\sc i\kern-.025em b}\kern-.08em
    T\kern-.1667em\lower.7ex\hbox{E}\kern-.125emX}}
\begin{document}

\title{An Effective Framework of Private Ethereum Blockchain Networks for Smart Grid  \\
\thanks{This  work  is  the  output  of  the  ASEAN IVO \url{http://www.nict.go.jp/en/asean_ivo/index.html} project ``Cyber-Attack Detection and Information Security for Industry 4.0'' and financially supported by NICT \url{http://www.nict.go.jp/en/index.html}. 
Correspondence: Nguyen Linh Trung.}
}

\makeatletter
\newcommand{\linebreakand}{%
  \end{@IEEEauthorhalign}
  \hfill\mbox{}\par
  \mbox{}\hfill\begin{@IEEEauthorhalign}
}
\makeatother

\author{
\IEEEauthorblockN{Do Hai Son\IEEEauthorrefmark{1}}
\IEEEauthorblockA{
dohaison1998@vnu.edu.vn}
\and
\IEEEauthorblockN{Tran Thi Thuy Quynh\IEEEauthorrefmark{1}}
\IEEEauthorblockA{ 
quynhttt@vnu.edu.vn}
\and
\IEEEauthorblockN{Tran Viet Khoa\IEEEauthorrefmark{1}\IEEEauthorrefmark{2}}
\IEEEauthorblockA{
khoatv.uet@vnu.edu.vn}
\linebreakand
\IEEEauthorblockN{Dinh Thai Hoang\IEEEauthorrefmark{2}}
\IEEEauthorblockA{
hoang.dinh@uts.edu.au}
\and
\IEEEauthorblockN{Nguyen Linh Trung\IEEEauthorrefmark{1}}
\IEEEauthorblockA{
linhtrung@vnu.edu.vn}
\and
\IEEEauthorblockN{Nguyen Viet Ha\IEEEauthorrefmark{1}}
\IEEEauthorblockA{hanv@vnu.edu.vn}
\linebreakand
\IEEEauthorblockN{\hspace{1cm}Dusit Niyato\IEEEauthorrefmark{3}}
\IEEEauthorblockA{
\hspace{1cm}dniyato@ntu.edu.sg}
\and
\IEEEauthorblockN{\hspace{1cm}Diep N. Nguyen\IEEEauthorrefmark{2}}
\IEEEauthorblockA{\hspace{1cm}diep.nguyen@uts.edu.au}
\and
\IEEEauthorblockN{Eryk Dutkiewicz\IEEEauthorrefmark{2}}
\IEEEauthorblockA{eryk.dutkiewicz@uts.edu.au}
\linebreakand
\IEEEauthorrefmark{1} AVITECH, VNU University of Engineering and Technology, Vietnam National University, Hanoi, Vietnam \\
\IEEEauthorrefmark{2} School of Electrical and Data Engineering, University of Technology Sydney, Australia \\
\IEEEauthorrefmark{3} Computer Science and Engineering, Nanyang Technological University, Nanyang, Singapore
}

\maketitle

\begin{abstract}
A smart grid is an important application in Industry 4.0 with a lot of new technologies and equipment working together. Hence, sensitive data stored in the smart grid is vulnerable to malicious modification and theft. This paper proposes a framework to build a smart grid based on a highly effective private Ethereum network. Our framework provides a real smart grid that includes modern hardware and a smart contract to secure data in the blockchain network. To obtain high throughput but a low uncle rate, the difficulty calculation method used in the mining process of the Ethereum consensus mechanism is modified to adapt to the practical smart grid setup. The performance in terms of throughput and latency are evaluated by simulation and verified by the real smart grid setup. The enhanced private Ethereum-based smart grid has significantly better performance than the public one. Moreover, this framework can be applied to any system used to store data in the Ethereum network.

\end{abstract}


\section{Introduction}
In Industry 4.0, a huge amount of data is created from smart applications. This leads to a serious security threat to many entities, including users, service providers, and network operators~\cite{attack}. Blockchain technology~\cite{bitcoin} has recently been introduced as a promising solution. It stores data on a distributed ledger and a consensus mechanism among the nodes to avoid the ledger being manipulated maliciously. 

In 2009, Bitcoin was introduced by Satoshi Nakamoto and hailed as a radical development in  money, being the first example of a digital asset. Since Bitcoin only intended to serve as a decentralized payment, the one's transaction will show up in a long time. That is why after Bitcoin, blockchain technologies are blooming with a famous project called ``Ethereum'' -- a decentralized platform. This technology not only possesses advantages of Bitcoin's technologies, i.e., decentralization, transparency, immutability, and security-and-privacy but also has some great improvements, i.e., smart contract and GHOST (Greedy Heaviest Observed Subtree) protocol~\cite{Sompolinsky}. It takes only $15$ seconds to confirm a new block, approximately $2.5$\% of Bitcoin~\cite{VitalikButerin2014}. As a result, the Ethereum network is applied in many impactful applications, such as smart agriculture~\cite{Mirabelli2020, ShyamalaDevi2019}, Internet-of-vehicles~\cite{Jabbar2020}, healthcare~\cite{DrewIvan2016, Hersh2016}.

\begin{figure*}[t]
	\centering     %%% not \center
	\includegraphics[width=\linewidth]{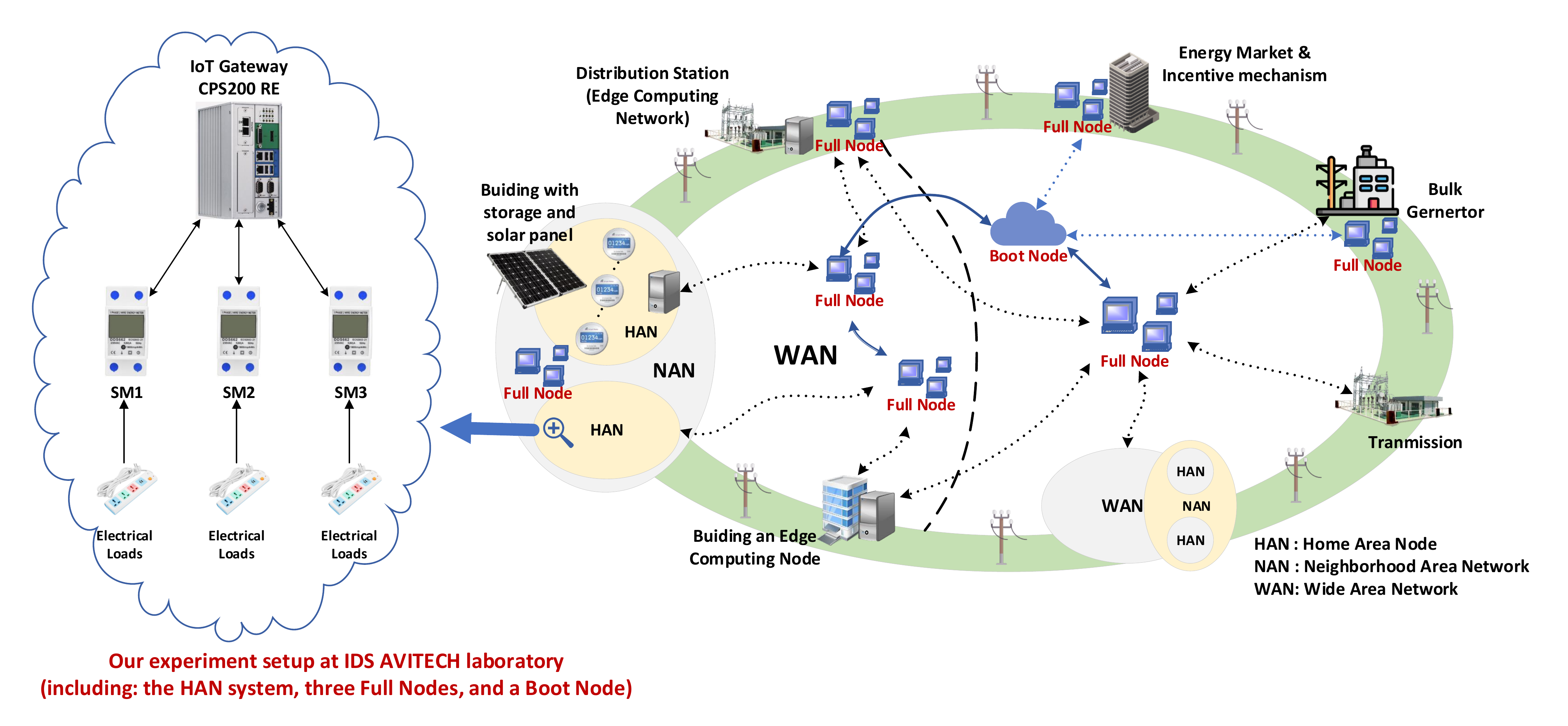}\\[-.5cm]
	\caption{Model of a smart grid in a smart city.}
	\label{fig:smartgrid}
\end{figure*}  

Industry 4.0 also sees the smart grid as an attractive application, which stores, manages, and exploits the electric system~\cite{El-Hawary2014}. A traditional grid is described in~\cite{Guan2019} with only one centralized server, controlled by the energy company, interacts with customers with symmetric or asymmetric schemes. A serious risk of the system is centralization, which leads to increasing latency and loss of data when cyber-attacks occur. The Ethereum network can be applied to overcome the risk~\cite{Zhuang2020, Huang2019, Gao2018}. A typical Ethereum-based smart grid is illustrated in Fig.~\ref{fig:smartgrid}. It is a potential system for the future electric network where users and energy markets are connected.

% privacy-sensitive information 

Several studies related to Ethereum-based smart grid are as follows. Zhuang {\em et al.}~\cite{Zhuang2020} reviewed the blockchain technology and showed the architecture and platform of a blockchain-based smart grid for cyber-security. Huang {\em et al.}~\cite{Huang2019} presented smart grid's protocols in theory and implemented a communication system using Sigfox devices, but did not apply the Ethereum network. Gao {\em et al.}~\cite{Gao2018} introduced a smart contract for their smart grid, but practical experiments were not investigated. Besides Ethereum, many studies applied the smart grid into other blockchain networks. \cite{Gai2019} used the Hyperledger Fabric 1.0 without optimization. \cite{Bansal2019} proposed a structure of their SmartChain framework, which used the concept of Proof-of-Time (PoT), instead of Proof-of-Work (PoW), for good computational and propagation time versus conventional blockchain network in simulation, but without verification. \cite{Kumari2020} proposed an energy trading (ET) framework without targeting a specific blockchain network, however, the performance was evaluated only in terms of the number of Hash functions to execute. Thus, it is unclear if this framework can be implemented in a real blockchain network. Generally, most of the studies reviewed above and others in the literature proposed system models or algorithms of the smart contracts for the smart grid or implementation of a testing system without the Ethereum network. This motivates us to focus, in this work, on developing a novel framework to implement a smart grid with secure data, and improve the efficiency of the private Ethereum network.

As shown in Fig.~\ref{fig:smartgrid}, a smart grid communication infrastructure can be separated into three layers: Home Area Node (HAN) which is the home electrical system, Neighborhood Area Network (NAN) which includes several HANs, Wide Area Network (WAN) which is a network of NANs. In this paper, a prototype of an Ethereum-based smart grid is implemented at the HAN layer. This prototype includes the essential components of a smart grid, e.g., smart meters and an IoT Gateway~\cite{IoT}. To secure data inside the blockchain network, encryption methods are required. However, asymmetry schemes would make users reveal their secret keys to the nodes~\cite{Raikwar2019}. Therefore, a symmetric pre-encryption technique and a simple smart contract are considered to prevent the encrypted data from being duplicated and to avoid revealing the secret~key.

The system is based on a private Ethereum network instead of a public Ethereum network. In a small-scale network, the mining time can be reduced while still ensuring security in the Ethereum network. By modifying the difficulty calculation method of the Ethereum consensus mechanism~\cite{wood2014}, the performance can be improved, with higher throughput, smaller latency, while keeping an uncle rate the same as that in the main Ethereum network. 
To experiment this method, we use BlockSim -- a recently proposed framework for blockchain systems by Alharby and van Moorsel~\cite{Alharby2020}. The input parameters for performance study by simulation are measured from the real system. Once we have obtained a suitable threshold of block interval in the consensus layer, the trade-off between latency and uncle rate, these parameters are then applied to the prototype to verify the system performance.

The main contribution of this paper is to propose an effective framework to build a private Ethereum network for a smart grid. Firstly, a practical Ethereum-based smart grid is deployed with essential hardware at the home electrical system. Secondly, a smart contract for authentication in a securely multi-devices system is proposed. At last, a method to improve the efficiency of an Ethereum-based smart grid setup in practical work with the support of numerical experiments.

%The remainder of the paper is organized as follows. In Section~\ref{sec:Ethereum}, the Ethereum blockchain technology is briefly described. Section~\ref{sec:proposedNet} presents the implementation of the proposed private Ethereum blockchain network for the smart grid and the proposed modification on the method of difficulty calculation for improving the performance. Section~\ref{sec:Experiments} performances improvement experiment on a real smart grid. Finally, Section~\ref{sec:Conclusion} concludes the paper.

\section{Overview of Ethereum Blockchain Technology}
\label{sec:Ethereum}
\subsection{Ethereum Blockchain Technology}

The Ethereum blockchain network was introduced by Vitalik Buterin in 2015.
% Different from Bitcoin, a new feature, call ``smart contract'' is added to the network. This makes Ethereum applicable for many purposes depending on the user's programming.
Typically, it can be divided into seven protocol layers: Storage, Data, Network, Protocol, Consensus, Contract, and Application. The Contract Layer and the GHOST protocol~\cite{Sompolinsky} in the Consensus Layer are greatly upgraded, in comparison with those in the Bitcoin network.
% ~\cite{Zhao2020}

The smart contract in the Contract Layer is the first highlight of Ethereum. A smart contract, being simply a piece of code running on Ethereum, can be built with the Solidity language. The Solidity Compiler compiles the smart contract into Bytecode and Application Binary Interface (ABI). Both of them are packaged into a transaction and deployed into the Ethereum network. Bytecode is an executable code on Ethereum Virtual Machine (EVM) and Contract ABI is an interface to interact with EVM Bytecode. 

% Web3~\cite{Web3.js2018} is a tool provided for users to interact with smart contracts.

GHOST is a PoW blockchain protocol like in Bitcoin, except for the way it resolves the correct blockchain. Instead of using the longest chain consensus rule in Bitcoin, GHOST follows the path of the sub-tree with the combined hardest proof of work/difficulty. The sub-tree is created because Ethereum allows us to reintroduce orphaned blocks to the chain as ``uncles''. These uncles in the chain allow the network to reduce the mining time while avoiding multiple forking of the ledger ($51$\%  attacks). 
% According to~\cite{VitalikButerin}, this protocol makes the block confirmation time of Ethereum significantly faster than Bitcoin's  (i.e., 15 seconds in the Ethereum versus 10 minutes in Bitcoin).
But the uncle has no role in data storage, so if the uncle rate is too high, it will lead to unnecessary storage effort.

\subsection{Types of Ethereum Nodes}

A node is a device/program that communicates with the Ethereum network, also known as a client. In this prototype, there are two node types of the Ethereum network.
% are shown in Fig.~\ref{fig:ETH_network}. 
A {\em full} node keeps a ledger, receives or broadcasts transactions to other nodes. Any full node can be used to confirm blocks and transactions and get rewards. In this case, it is also called ``miner''.
A {\em boot} node keeps Ethereum Node Records (ENR) of many full nodes and is not responsible for keeping the ledger, mining, or broadcasting transactions. Any node that connects with a boot node would discover peers in the network.

% Besides, a Web client is connected to the Ethereum network to read and show data via a dashboard interface as shown in Fig.~\ref{fig:private}.
% \begin{figure}[t]
% 	\centering     %%% not \center
% 	\includegraphics[width=0.45\textwidth]{figure/private.png}
% 	\caption{Dashboard of the smart grid using private testnet.}
% 	\label{fig:private}
% \end{figure}  
% \begin{figure}[t!]
% 	\centering     %%% not \center
% 	\includegraphics[width=0.45\textwidth]{figure/ETH_network.pdf}
% 	\caption{The Ethereum Network model.}
% 	\label{fig:ETH_network}
% \end{figure}

\section{Proposed private Ethereum network for\\ smart grid}
\label{sec:proposedNet}
\subsection{Private Ethereum Network and Hardware Implementation}
% \label{sec:3.2}
\label{subsec:Sys}
At the present, 1 Ether (ETH) is approximately $2000$~USD. If the smart grid is deployed in the public Ethereum network, at least transaction fees will be charged. There are two ways to deploy the system and send transaction fees in the Ethereum network; one uses Ethereum test-nets and the other creates an own private network.

In the former, test-nets are for free by given ETH coins in some vaults, and the benefit of this way is that we do not necessarily run our miners. However, the ETH coins obtained from vaults are limited. So for a large-scale system, Ethereum test-nets are not compatible. 

In the latter, a private Ethereum network is preferred because it can overcome the disadvantage of the former. To deploy a private network, the Ethereum developer team provides a powerful open-source software named Go-ethereum~\cite{Ethereum} (Geth). The software includes a number of functions to make private nodes, make boot nodes, create new accounts, run full nodes, and so on. Geth {v1.10.4} is used in our setup. The private Ethereum network,  as shown in Fig.~\ref{fig:miner}, consists of three full nodes which are personal computers with processor Intel® Core™ i7-4800MQ @2.7 GHz, RAM of 16 GB. The network has been setup by a Cisco switch Catalyst 2950 with 100 Mbps bandwidth.

\begin{figure}[t]
	\centering     %%% not \center
	\includegraphics[width=\linewidth]{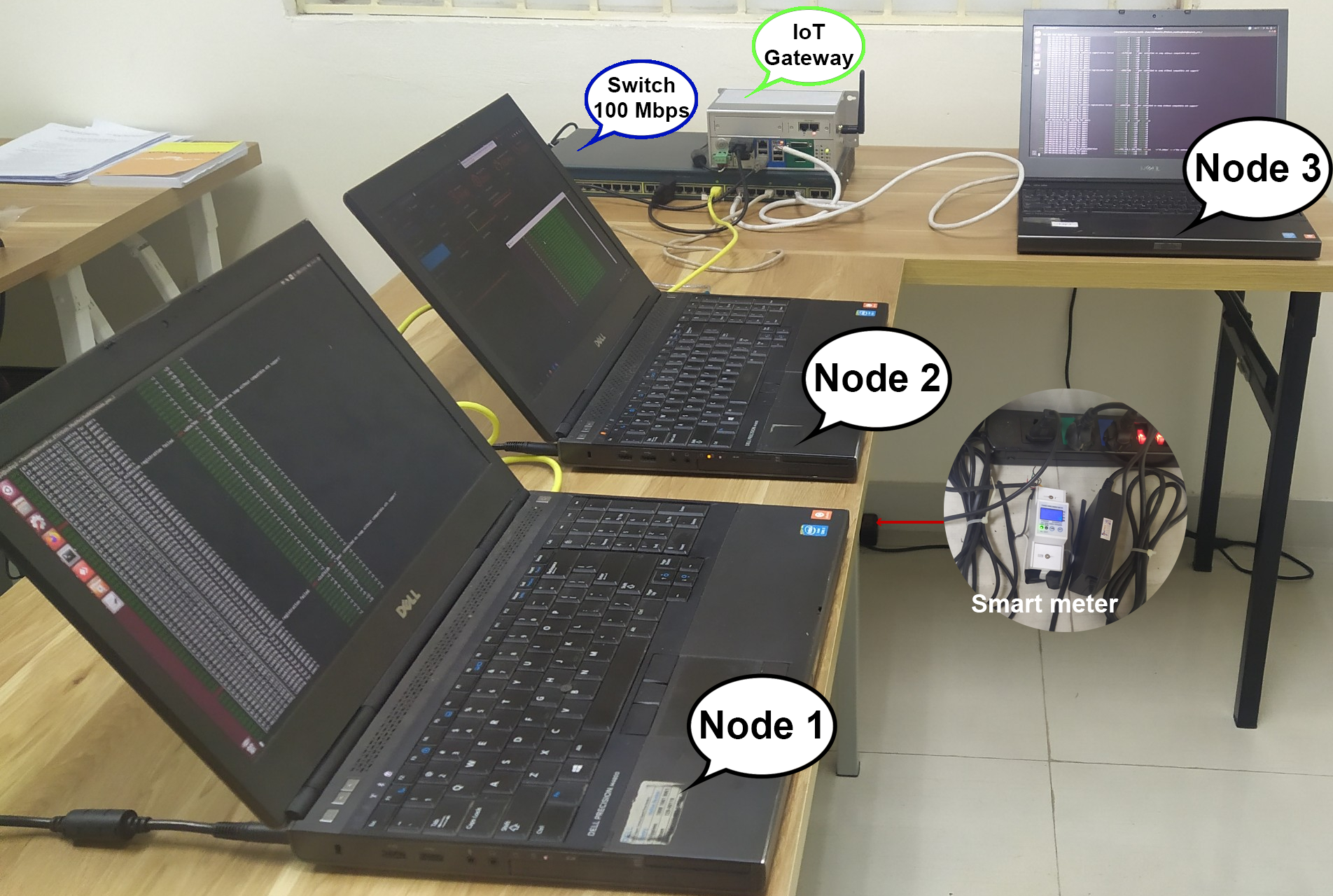}
	\caption{Our experiment system.}
	\label{fig:miner}
\end{figure}

As shown in Fig.~\ref{fig:smartgrid}, the proposed prototype of the HAN layer includes the essential components, i.e., electrical loads, smart meters, and an IoT gateway CPS 200RE. 

XTM35SC is the next generation of electricity meters (smart meters). It measures how much electricity has been used, and displays information on a handy in-home display. Furthermore, data collected from the smart meter can be exploited by other IoT devices which use the Modbus-RTU protocol. In the system, data collected from the smart meter will be exploited, decoded, encrypted, and transmitted through an IoT gateway. For simplicity, only consumed energy data will be collected.

CPS 200RE is an edge IoT gateway. It is fully integrated with Fieldbus accessibility, Modbus TCP/RTU, PROFINET® or EtherNet/IP™, and so on, for extremely easy deployment of both centralized/decentralized field data implementation in the automation process. The IoT gateway is responsible for the collection of device identification, collection time, and value of consumed energy.

\subsection{Security Enhanced Smart Contract}

After collecting raw data from smart meters, the gateway encrypts the data to avoid malicious tapping of the data. This work is necessary because the mechanism of the blockchain makes all data public, requiring pre-encryption before transmission. Both symmetry and asymmetry schemes are considered. But in the blockchain, a classical symmetry scheme named AES-256-CTR~\cite{Dworkin2001} is used in the system because asymmetry schemes would make a user reveal his/her $key_{private}$ to Ethereum nodes~\cite{Raikwar2019}. In the blockchain network, a pair of $key_{pub}$ and $key_{pri}$ are provided when a user creates a new account. In this paper, the $key_{private}$ used for AES-256-CTR is the same as the $key_{pri}$. At this stage, the privacy of the raw data is guaranteed.

The smart contract is given by Algorithm~\hyperref[alg:smartcontract]{1} in which the inputs are encrypted identification/collection time/value of consumed energy $\mathcal{F}($\ldots$, key_{pri})$ and the outputs are stored inputs, saving the $key_{pub}$ of the user account that has deployed the smart contract as the primary account. This account has permission to add/remove other accounts from the account list, allowing the added/removed account to push data in smart contract or not. 
% Then, the contract is deployed in Remix - Ethereum IDE, and transaction cost per function is captured and shown in Table~\hyperref[txgas]{I}. 
The data flow of every smart meter to the private Ethereum network is summarized in Fig~\ref{fig:data_flow}. 
{
\begin{algorithm}[t]
	\renewcommand{\thealgorithm}{}
	% 	\floatname{algorithm}{Smart Contract:}
	\caption{\textbf{1} \\
		\textbf{Input}: msg.sender (address sending transaction) \\
		\hspace*{0.9cm} \_id $\longleftarrow \mathcal{F}(\_id, key_{pri})$     \\
		\hspace*{0.9cm} \_time $\longleftarrow \mathcal{F}(\_time, key_{pri})$     \\
		\hspace*{0.9cm} \_value $\longleftarrow \mathcal{F}(\_value, key_{pri})$     \\
		\textbf{Output}: push data to the Ethereum network}
	\label{alg:smartcontract}
	\begin{algorithmic}[1]
		\item \textbf{Class} SmartFac $\;$ \{
		\item \hspace{0.25cm} init\_addr $\longleftarrow$ None
		\item \hspace{0.25cm} total\_of\_reco $\longleftarrow$ 0
		\item \hspace{0.25cm} struct Reco \{ id, time, value \}
		\item \hspace{0.25cm} reco[uint][Reco]
		\item \hspace{0.25cm} trusted\_acc[address][bool]
		
		\item \hspace{0.25cm} \textbf{Function} constructor(  ) \{
		\item \hspace{0.5cm} trusted\_acc[msg.sender] $\longleftarrow$ true
		\item \hspace{0.5cm} init\_addr $\longleftarrow$ msg.sender
		\item \hspace{0.25cm}   \}
		
		\item \hspace{0.25cm} \textbf{Function} add\_acc( \_addr ) \{
		\item \hspace{0.5cm} \textbf{if} msg.sender $\neq$ init\_addr
		\item \hspace{0.75cm} \textbf{return} Error	
		\item \hspace{0.5cm} trusted\_acc[\_addr] $\longleftarrow$ true	
		\item \hspace{0.25cm}   \}		
		
		\item \hspace{0.25cm} \textbf{Function} rm\_acc( \_addr ) \{
		\item \hspace{0.5cm} \textbf{if} msg.sender $\neq$ init\_addr $\|$ \_addr == init\_addr
		\item \hspace{0.75cm} \textbf{return} Error	
		\item \hspace{0.5cm} \textbf{delete} trusted\_acc[\_addr]	
		\item \hspace{0.25cm}   \}
		
		\item \hspace{0.25cm} \textbf{Event} added\_reco(addr, \_id, \_time, \_value)
		
		\item \hspace{0.25cm} \textbf{Function} new\_reco( \_id, \_time, \_value ) \{
		\item \hspace{0.5cm} \textbf{if} trusted\_acc[msg.sender] $\neq$ true
		\item \hspace{0.75cm} \textbf{return} Error		
		\item \hspace{0.5cm} total\_of\_reco $\longleftarrow$ total\_of\_reco + 1
		\item \hspace{0.5cm} reco[total\_of\_reco] $\longleftarrow$ Reco \{\_id,\_time,\_value\}
		\item \hspace{0.5cm} \textbf{emit added\_reco}(msg.sender, \_id, \_time, \_value)
		\item \hspace{0.25cm}   \}			
		\item \}
	\end{algorithmic}
\end{algorithm}
}
\begin{figure*}[t]
	\centering     %%% not \center
	\includegraphics[width=1\linewidth]{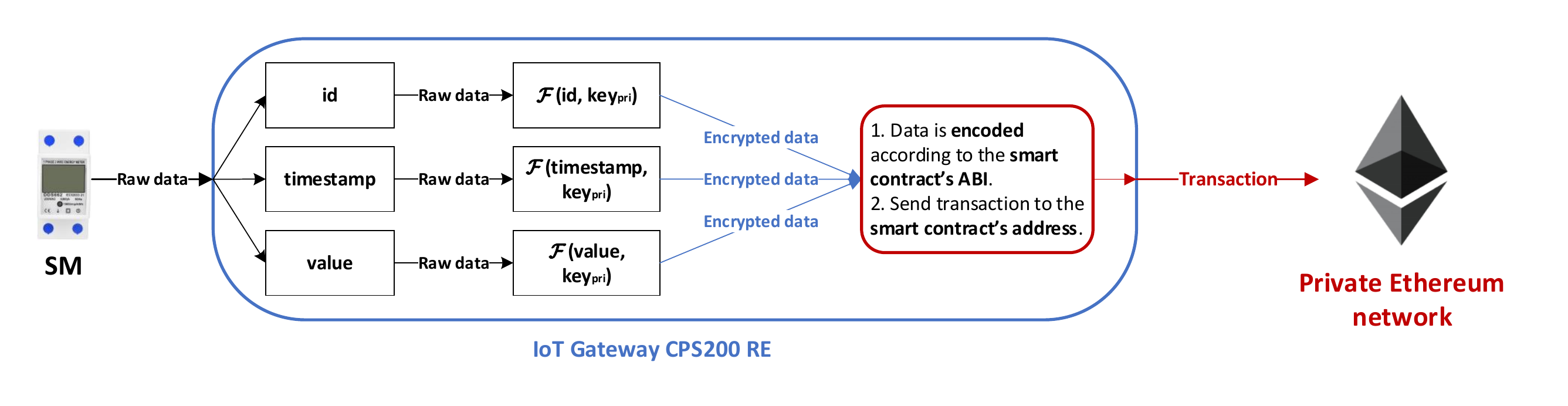}\\[-.5cm]
	\caption{Data flow from a smart meter to the private Ethereum network.}
	\label{fig:data_flow}
\end{figure*}

% \begin{table}[tb]
% 	\centering
% 	\label{txgas}
% 	\caption{Transaction cost per function in the smart contract.}
% 	\begin{tabular}{|c|r|}
% 		\hline
% 		\multicolumn{1}{|c|}{\textbf{Function}} & \multicolumn{1}{c|}{\textbf{Transaction cost (gas)}} \\ \hline
% 		Deploy smart contract & 573 581 \\ \hline
% 		add\_acc              & 26 960    \\ \hline
% 		rm\_acc               & 29 197    \\ \hline
% 		new\_reco             & 94 408   \\ \hline
% 	\end{tabular}
% \end{table}

\subsection{Performance Improvement via Throughput and Latency}

The latency of the original Ethereum network is more than 12 seconds because of global scalability~\cite{VitalikButerin2014}. This is not suitable for low-latency applications. When applied to a smart grid, with smaller scalability, the system can be improved to obtain higher throughput and smaller block intervals ($T$). 

Based on the analysis of $10,000$ consecutive blocks in the Bitcoin network, on average, the transmission time of a block which has just been produced from a miner to $50$\% and $95$\% of all nodes are $6.5$ seconds and $40$ seconds respectively, and the mean delay is around $12.6$ seconds~\cite{blocktime}. For a private Ethereum network with small scalability, the mining time can be reduced subject to block propagation, which is measured in this private Ethereum network. 

Firstly, we consider the case when the public Ethereum network keeps the stable block interval described in the Ethereum yellow paper~\cite{wood2014}:

\begin{equation}
    \label{eq:1}
    D_i = 
    \begin{cases}
        D_0 = 131072, & \text{if } i = 0, \\ 
        \max(D_0, P_D + x \times \zeta + \epsilon), & \text{otherwise},
    \end{cases}
\end{equation}
with
\begin{align}
    x &= \left \lfloor \frac{P_D}{2048} \right \rfloor,\\
    \label{eq:3}
    \zeta &= \max 
            \left\{y - \left \lfloor \frac{T}{9} \right \rfloor, -99\right\},\\
    y &= 
        \begin{cases}
            1, & \text{if } \left \| P_U \right \| =0,\\ 
            2, & \text{otherwise},
        \end{cases}\\
    \epsilon &= \left \lfloor 2^{\left \lfloor \max (i - 5000000, 0) \div 100000 \right \rfloor - 2} \right \rfloor,
\end{align}
where $D$ is ``difficulty'' which is a scalar value corresponding to the difficulty level of this block, $P$ is the parent of this block, $P_D$, $P_s$ and $P_U$ are ``difficulty'', ``time\_stamp'', and ``the number of uncles'' of $P$, respectively, $\left \lfloor \cdot \right \rfloor$ denotes the the integer division operator, the index $i$ indicates the current block number, T is the block interval, given by $T = \text{current\_block\_time\_stamp} - P_s$.

\begin{table*}
	\centering
	\caption{The parameters in the network are pre-setup based on the measured and our experiments.}
	\label{tab:par}
	\begin{tabular}{|l|l|l|}
    \hline
    \multicolumn{1}{|c|}{\textbf{Parameters}}            & \multicolumn{1}{c|}{\textbf{Values}} & \multicolumn{1}{c|}{\textbf{Notes}}                       \\ \hline
    Block gas limit                           & 15,000,000 gas             & Same as main Ethereum network at June-6-2021   \\ \hline
    Average transaction size                   & 0.759808 kB                & Based on our smart contract                                  \\ \hline
    Average block size                        & 60 kB                      & Same as main Ethereum network at June-6-2021   \\ \hline
    Average block propagation delay           & 0.25 seconds                     & Based on our experiment measured \\ \hline
    Sync mode                                  & light                      & Just sync block header                         \\ \hline
    Number of transactions created per second & 100 transactions per second                  &                                                 \\ \hline
    Number of node                            & 3 miners                   & Each miner has 33.33\% of the total computing power    \\ \hline
    \end{tabular}
\end{table*}

We focus on the case in which $\left \| P_U \right \| = 0$. By this constraint, we consider the effect of the propagation time to $T$. Simultaneously, when the number of uncles is reduced, the size of the ledger is decreased while still keeping all transactions. Moreover, reducing the number of uncles in the ledger also avoids ``selfish~mining'' in the network when some miners are simply to mine uncles instead of blocks extending the best chain~\cite{uncle}.

We can observe that the Ethereum consensus does not vary $T$ directly, but indirectly through $D$ and a threshold $\lambda$ in replacement of value $9$ in~\eqref{eq:3}. In detail, Eq.~\eqref{eq:1} depends on Eq.~\eqref{eq:3}, so unless the current block is the first one (genesis block), if $T < \lambda$ then $D$ is adjusted upwards by $(x + \epsilon)$, if $\lambda \le T < 2\lambda$ then $D$ is unchanged, and if $T \ge 2\lambda$  then $D$ is adjusted downwards proportional to the timestamp difference by from $(-x + \epsilon)$ to $(-99 \times x + \epsilon)$.
It can be seen that $T$ is always desired between $\lambda$ and $2\lambda$ seconds, subject to $\lambda$ in \eqref{eq:3}. It should be well noted that this threshold is rooted from~\cite{blocktime}.

However, in a private Ethereum network of limited size, $T$ can be reduced while ensuring that the block propagates through 95\% of the nodes. This leads to our modification in the consensus mechanism to improve the block interval. We propose to modify the source code of Geth~\cite{Ethereum} by restoring the use of the threshold $\lambda$ in~\cite{blocktime} as in
\begin{equation}
    \label{eq:6}
    \zeta = \max 
    \left\{y - \left \lfloor \frac{T}{\lambda} \right \rfloor, -99\right\}
\end{equation}
instead of fixing it to $9$ as in~\eqref{eq:3}, and obtain the value of $\lambda$ based on the practical smart grid setup. 

\section{Experiments}
\label{sec:Experiments}
\subsection{Performance Study by Simulation}
\label{subsec:Simu}
Recently, a simulation framework called BlockSim~\cite{Alharby2020} is introduced. It is used to evaluate a Bitcoin/Ethereum blockchain network depending on input parameters. This paper focuses on evaluating three parameters, i.e., block interval, throughput (transactions per second), and uncle rate. The others based on either the main (public) Ethereum network or measured data from the prototype, and are given in Table~\ref{tab:par}.

The simulation results are shown in Fig.~\ref{fig:simulation}. The line with marker of triangle and that of diamond present uncle rate and throughput versus block interval, respectively. The results are averaged over 100 experiments. 

Generally, the throughput and the uncle rate both decrease as the block interval time increases. The system has maximum performance with throughput of more than 80 transactions per second (tx/s) and block interval of less than 2 seconds. However, $7$-$18$\% of uncle rate is not good because of increasing storage. Comparing with the main Ethereum network of $4.81$\% on average~\cite{unclerate}, the proposed network has $T = 3$ seconds. At this value of block interval, throughput and uncle rate are $77.72$~tx/s and $4.88$\%, compared with $29.95$ tx/s and $1.47$\% at $T=12$, respectively.

Besides, other simulation results of a main Ethereum network~\cite{blocktime} with 3 full nodes are two scatter points as shown in Fig.~\ref{fig:simulation} with throughput and uncle rate are $14.05$ tx/s and $17.48$\%, respectively. This  throughput is suitable with the real main Ethereum network.

\begin{figure}[t]
	\centering     %%% not \center
	\includegraphics[width=\linewidth]{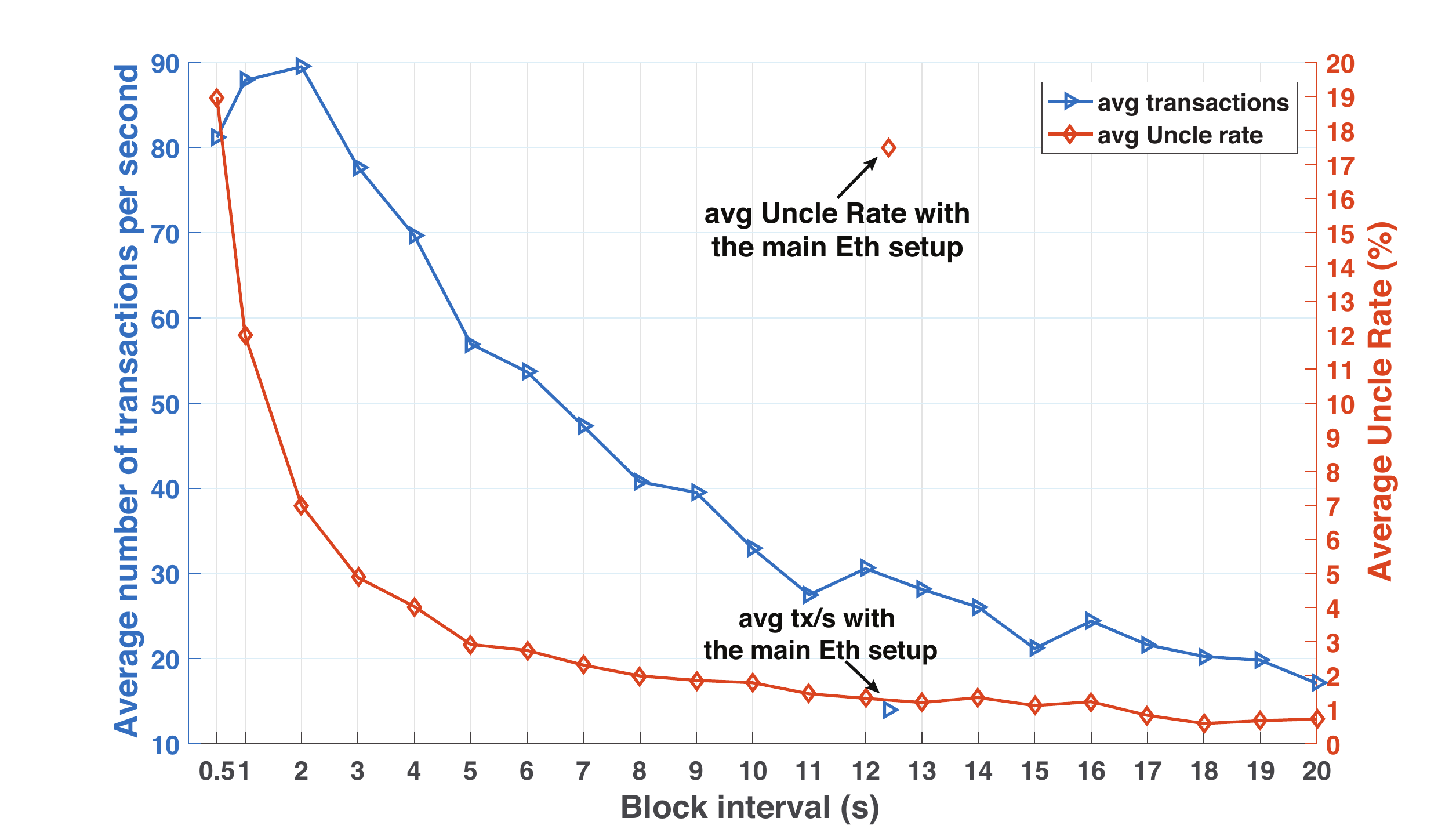}
	\caption{Performance by simulation: Throughput and uncle rate versus block interval.}
	\label{fig:simulation}
\end{figure}

% {\color{red}Need to inform the value of $\lambda$ for this configuration, in Table II}.

\subsection{Verification by Real Data}

From the experiment by simulation, to obtain the block interval of approximately $3$ seconds and the uncle rate of approximately 4.81\%, as shown in Section~\ref{subsec:Simu}, the threshold $\lambda$ in Eq.~\eqref{eq:6} is found to be $3$. 

\begin{figure}[t]
	\centering     %%% not \center
	\includegraphics[width=0.47\textwidth]{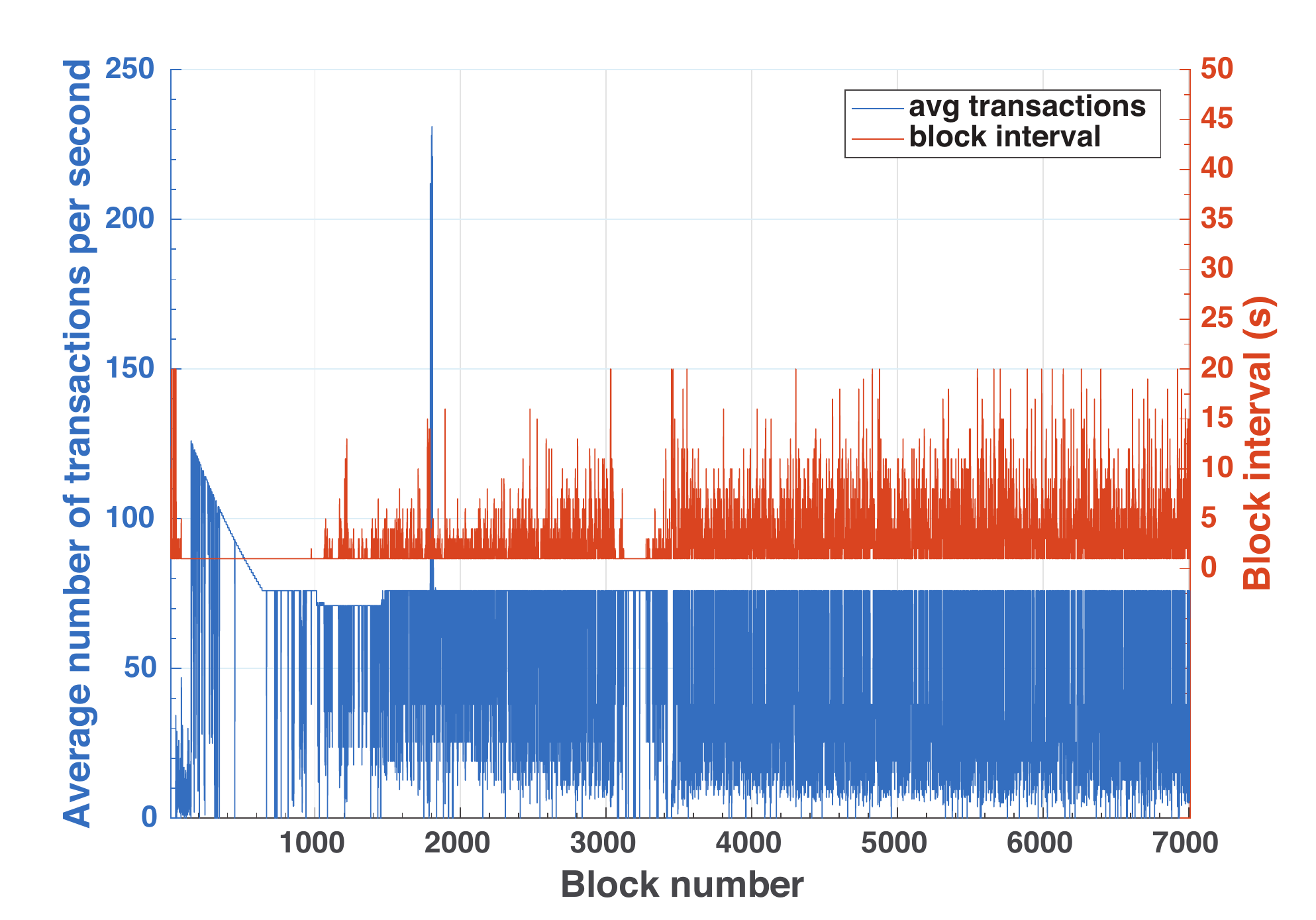}
	\caption{Real experiment: Throughput and block interval in $7,000$ blocks at $\lambda$ = 3.}
	\label{fig:real}
\end{figure} 

The experimental results from the real prototype are presented in Fig.~\ref{fig:real} with $7,000$ consecutive blocks. The top line and bottom line show the average number of transactions per second in a block time  and the block interval, respectively. Here, because the mining process depends on probability, so in this figure, we can see that both are not stable at a specific value. The average values of both on $7,000$ blocks, compared to that of the main Ethereum network (found at \url{www.etherscan.io/chart}), are shown in Table~\ref{tab:3}. According to the real experiment, the proposed framework can handle $50.08$~tx/s and $2.7$ seconds between two blocks while those of the main Ethereum network are 16.25~tx/s and 13.48~seconds. Moreover, the obtained uncle rate is 3.03\%, which is smaller than that of the main Ethereum network.

The throughput of the real experiment is not as good as that of the simulated experiment by BlockSim but better than that of the main Ethereum network. While the latency is much smaller than that of the main Ethereum network. It is clear that the proposal of changing the consensus mechanism did give better performance.

\begin{table}
\centering
\caption{Experiment results versus the main Ethereum network}
\label{tab:3}
\begin{tabular}{|l|l|l|} 
\hline
\multicolumn{1}{|c|}{\textbf{Parameters}}                           & \multicolumn{1}{c|}{\begin{tabular}[c]{@{}c@{}}\textbf{Avg values}\\\textbf{ in the private Eth}\end{tabular}} & \multicolumn{1}{c|}{\begin{tabular}[c]{@{}c@{}}\textbf{Avg values}\\\textbf{ in the main Eth}\end{tabular}}  \\ 
\hline
Transactions per second                                            & 50.08 tx/s                                                                                                    & 16.25 tx/s                                                                                                  \\ 
\hline
Uncle Rate                                                         & 3.03\%                                                                                                       & 4.81\%                                                                                                     \\ 
\hline
Block interval                                                     & 2.7 seconds                                                                                                        & 13.48 seconds                                                                                                     \\ 

\hline
\end{tabular}
\end{table}

However, decreasing $T$ may cause risky scenarios, for example, a significantly faster miner joins the network and takes all mining jobs. It can lead to a $51$\% attack by confirming dishonest transactions. In a private network, there are a few technical methods provided by Geth to solve this problem, e.g., whitelist IP, set maxpeers; more details in~\cite{Ethereum}.

\section{Conclusion}
\label{sec:Conclusion}
% The work aims to build a highly-effective together with a prototype of a private blockchain network in order to ensure the privacy and at the same time enhance system performance for smart grid. Through the real experimental results, we show that our proposed system can obtain throughput of 50.08 tx/s and latency of 2.7 seconds at 3.03\% Uncle Rate. These results clearly show that our proposed framework can outperform the current settings of current public Ethereum network. Though, the team will need finding solutions to overcome decreasing of dispersibility of the network in the next phase.

We have proposed an effective framework to build a private Ethereum network for smart grid with an own private Ethereum network and essential hardware of a smart grid at the home electrical system. The AES-256-CTR standard is applied to pre-encrypt raw data and a smart contract for authentication has been proposed. Then, we have shown how to improve the efficiency of a practical smart grid setup and our verification system can obtain throughput of $50.08$ tx/s and latency of $2.7$~seconds at an uncle rate of $3.03$\%. These results clearly show that our proposed framework can outperform the original setup for a private Ethereum network. Moreover, this framework can be applied to any system used to store data in the Ethereum network with any scale. In the future, other factors, e.g., the number of nodes, transaction size, will be investigated to fully evaluate a private Ethereum network.

\balance

\end{document}